\documentclass[aps,reprint,superscriptaddress,floatfix,showpacs]{revtex4-1}
\usepackage{xcolor}
\usepackage{cancel}

\usepackage{float}
\usepackage{gensymb}
\usepackage{lipsum}
\usepackage{amsmath}
\usepackage{amssymb}
\usepackage{amssymb}
\usepackage{empheq} 
\usepackage{braket}

\usepackage{bm}
\usepackage{graphicx}
\usepackage{hyperref}
\usepackage{indentfirst}
\usepackage{siunitx}
\usepackage{soul} 
\hypersetup{
    colorlinks,
    linkcolor={blue!50!blue},
    citecolor={blue!50!blue},
    urlcolor={blue!
    80!black}
}

\begin{document}

\title{Mid infrared polarization engineering via sub-wavelength biaxial hyperbolic van der Waals crystals}
\author{Saurabh Dixit}
\author{Nihar Ranjan Sahoo}
\author{Abhishek Mall}
\author{Anshuman Kumar}
\email{anshuman.kumar@iitb.ac.in}
\affiliation{Laboratory of Optics of Quantum Materials, Department of Physics, IIT Bombay, Mumbai - 400076, India}
\date{\today}
\keywords{Natural hyperbolic metamaterials, Van der Waal material, 2D thin film, Polarizer, Waveplate}

\begin{abstract}
Recently, in-plane biaxial hyperbolicity has been observed in $\alpha$-MoO${_3}$ --a van der Waal crystal-- in the mid-infrared frequency regime. Here, we present a comprehensive theoretical analysis of thin film $\alpha$-MoO${_3}$ for application to two mid-IR photonic devices -- a polarizer and a waveplate. We show the possibility of a significant reduction in the device footprint while maintaining an enormous extinction ratio from $\alpha$-MoO${_3}$ based polarizers in comparison with that of conventional polarizers. Secondly, we carry out device optimization of $\alpha$-MoO${_3}$ thin-film based waveplates with subwavelength thickness. We explain our results using natural in-plane hyperbolicity of $\alpha$-MoO${_3}$ via analytical and full wave electromagnetic simulations. This work will build a foundation for miniaturization of mid-infrared photonic devices by exploiting the optical anisotropy of $\alpha$-MoO${_3}$.  
\end{abstract}

\maketitle


The mid-infrared (mid-IR) region of the electromagnetic spectrum is relevant for a number of applications\cite{Pile2012} in the areas of medical diagnostics\cite{Xue2019,Oh2018}, thermal imaging\cite{Park2018}, molecular sensing\cite{Martnez2016}, free-space optical communication\cite{Hao2017}, among others. Therefore, there is a particular interest in the development of mid-IR components such as sources, detectors, and other opto-electronic components\cite{Lin2017,Gansel2009}. One of the critical challenges for the development of mid-IR technologies is the miniaturization of optical components\cite{Law2013,Fang2019}. In particular, optical components for long and very long wavelength IR range (i.e., 8$\mu$m - 20$\mu$m) are relatively less developed than the mid wavelength (i.e., 3$\mu$m - 8$\mu$m). For instance, conventional wire grid polarizers available in the long-wavelength IR (LWIR) region, are bulky and typically exhibit around 70\% transmission efficiency and 20 dB of ER\cite{polarizer}. Besides, phase retarders for $\lambda$/2 and $\lambda$/4 are mostly available for mid-IR wavelength range only with the thickness in millimeters. To this end, the newly discovered van der Waals (vdW) materials can enable a new class of small footprint mid-IR photonic components as well as their easy integration with conventional platforms through van der Waals integration\cite{Caldwell2016,Kim2019,Guo2017,Liu2019,Fang2019}.\par

 Hyperbolic metamaterials (HMM) are a class of anisotropic materials that have opposite signs of principle components of the real part of dielectric permittivity tensor\cite{Poddubny2013}. Hence, HMM behaves like a metal in one crystal direction and dielectric in the other. This property has enabled several exciting applications such as hyperlens, negative refraction, thermal emission engineering, and others\cite{Guo2020}. Recently, it has been experimentally demonstrated that orthorhombic molybdenum trioxide  ($\alpha$-MoO${_3}$) possesses biaxial in-plane hyperbolicity in the mid-IR\cite{Ma2018,Zheng2018,Zheng2019}. Orthorhombic crystal structure of $\alpha$-MoO${_3}$, are formed by distorted MoO${_6}$ octahedra and bounded with vdW forces with three different lattice constants ($a, b,$ and $c$)\cite{Ma2018}. The in-plane lattice constants of $\alpha$-MoO${_3}$ have a difference of 7\%, resulting in a strong in-plane optical anisotropy\cite{lvarezPrez2020}. Thus, this intrinsic giant in-plane anisotropy of $\alpha$-MoO${_3}$ can enable highly efficient and small footprint mid-IR optical components such as polarizers and waveplates without the need for lithographic patterning\cite{Xia2019,Nordin1999}.
 
In this letter, we examine the optical response of $\alpha$-MoO${_3}$ thin films as a function of thickness and frequency for the optimization of mid-IR photonic device (schematically shown in Fig.~1(a)). To evaluate the optical response and performance metrics of thin-film $\alpha$-MoO${_3}$ on substrate based polarizer and waveplate, we used the well-known transfer matrix method (TMM) (see Sec.~S1 of supporting information). We further corroborate the results of our analytical model via finite difference frequency domain (FDFD) numerical simulation using \textsc{comsol multiphysics}. The optical response of $\alpha-$MoO$_{3}$ in the mid-IR spectral region is dominated by the optical phonons and its dielectric permittivity tensor has been shown to follow a Lorentz model\cite{Zheng2019}:
\begin{equation}
    \varepsilon_{j} = \varepsilon_j^{\infty} \left(1 + \frac{(\omega_j^{LO})^2 - (\omega_j^{TO})^2}{(\omega_j^{TO})^2 - \omega^2 - i\omega\Gamma_j}\right)
\end{equation}
\begin{figure}
    \centering
    \includegraphics[width=\columnwidth]{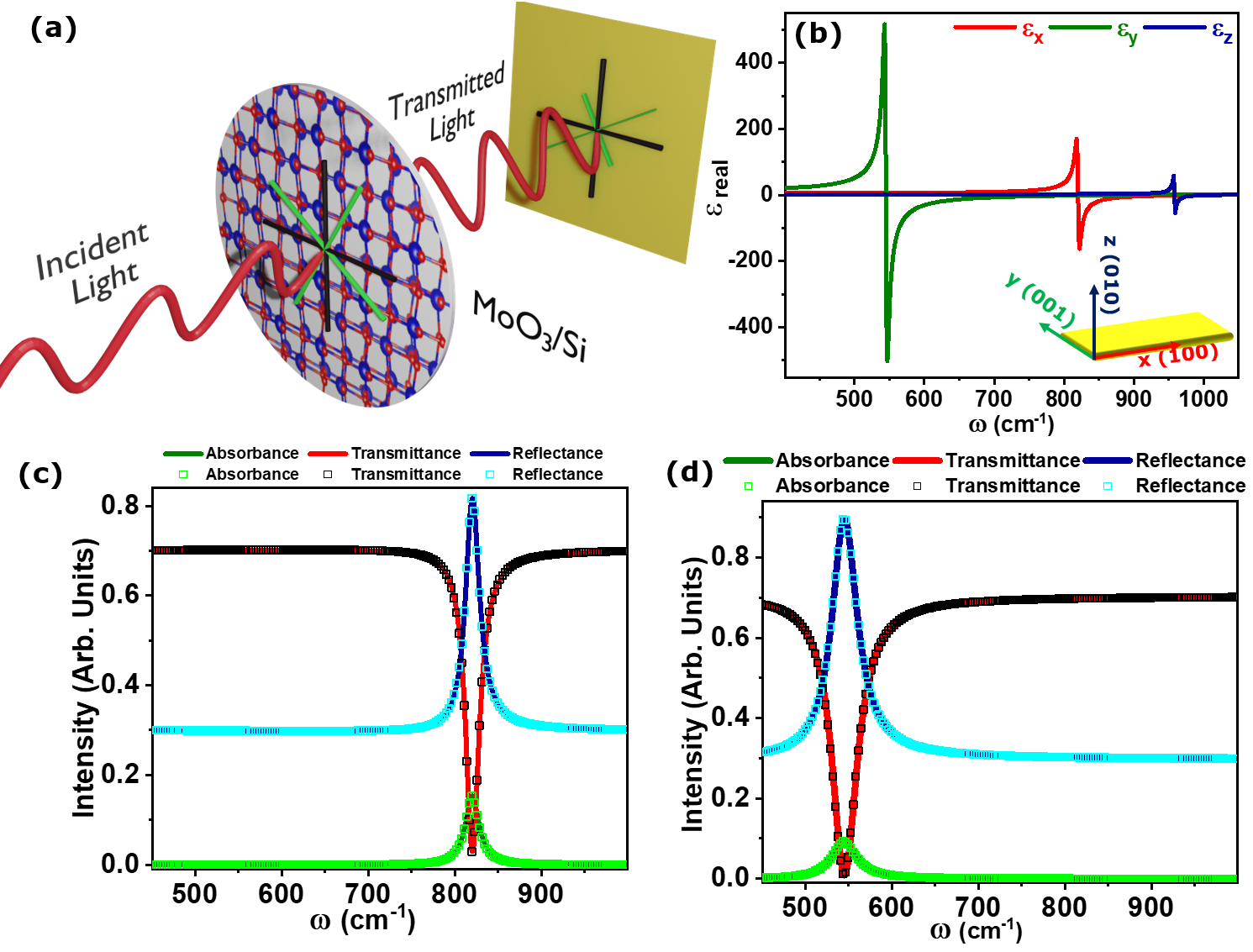}
    \caption{(a) Schematic diagram of $\alpha-$MoO$_{3}$ on silicon substrate based waveplate. (b) Real part of dielectric permittivity along $x, y$ and $z$ direction of $\alpha-$MoO$_{3}$ revealing the natural hyperbolicity of $\alpha-$MoO$_{3}$ in the frequency range of 545 cm$^{-1}$ to 1010 cm$^{-1}$. (c) and (d) represent the optical response of 100 nm thin film of $\alpha-$MoO$_{3}$ for $x-$ and $y-$ polarized light respectively. The scatter plot in both spectra correspond to numerical simulation results for the optical responses of $\alpha-$MoO$_{3}$.}
\end{figure}
where $j$ represents principle axes direction (i.e. $x, y$ and $z$) in the $\alpha-$MoO$_{3}$ crystal. The directions $x, y$ and $z$ correspond to [100], [001] and [010] crystalline directions respectively.  Here $\varepsilon_{j}$ is the principle component of dielectric tensor whereas $\varepsilon_j^{\infty}$, $\omega_j^{LO}$ and $\omega_j^{TO}$ represent high-frequency dielectric constant, frequency of longitudinal optical (LO) and transverse optical (TO) phonons respectively. Lastly, $\Gamma_{j}$ represents the line-width of the oscillations in the respective direction. Real part of the principle components of the dielectric tensor is plotted as a function of frequency in Fig.~1(b). The dielectric tensor reveals hyperbolicity of $\alpha-$MoO$_{3}$ from 10 $\mu$m to 18 $\mu$m. The material exhibits three Reststrahalen (RS) bands -- the spectral region between LO and TO phonons -- in the range of 545 cm$^{-1}$ to 1010 cm$^{-1}$. Bands 1--3 lie in the range of 545 cm$^{-1}$ to 850 cm$^{-1}$, 820 cm$^{-1}$ to  973 cm$^{-1}$ and  958 cm$^{-1}$ to  1010 cm$^{-1}$, where dielectric constant is negative along $y, x$ and $z$ directions (i.e. along [001] [100] and [010] directions) respectively. This property fulfills the fundamental criterion of desirable material for mid-IR optical components, as explained later in this paper.

Using the above dielectric permittivity tensor, we develop our TMM model for assessing the optical response of $\alpha-$MoO$_{3}$ as shown schematically in Fig.~1(a). Optical response for $x$-polarized and $y$-polarized components of incident light on a 100 nm thin film of $\alpha-$MoO$_{3}$ has been shown in the Fig.~1(c)-(d) respectively. A sharp dip is observed in the transmittance spectrum of $x$-polarized light (Fig.~1(c)) around 820 cm$^{-1}$. It is a manifestation of TO phonons of $\alpha$-MoO$_{3}$ along [100] direction due to which dielectric permittivity along that direction becomes negative. Hence, it reflects the light at this particular frequency, as shown by the reflectance of $x$-polarized light in Fig.~1(c). A similar phenomenon is observed in the optical responses of $y$-polarized light where $\alpha$-MoO$_{3}$ has high reflectance at the frequency of TO phonons along [001] direction (as shown in Fig.~1(d)). A small absorption by $\alpha-$MoO$_{3}$ thin film is also observed, which corresponds to optical losses in the dielectric material. Our FDFD simulation results are found to be in excellent agreement with the results from our TMM model. From this optical response, it is noticeable that a thin film of $\alpha$-MoO$_{3}$ can reflect light with one state of polarization and pass the light with the second state of polarization at TO phonon frequency. 
\begin{figure}
    \centering
    \includegraphics[width=\columnwidth]{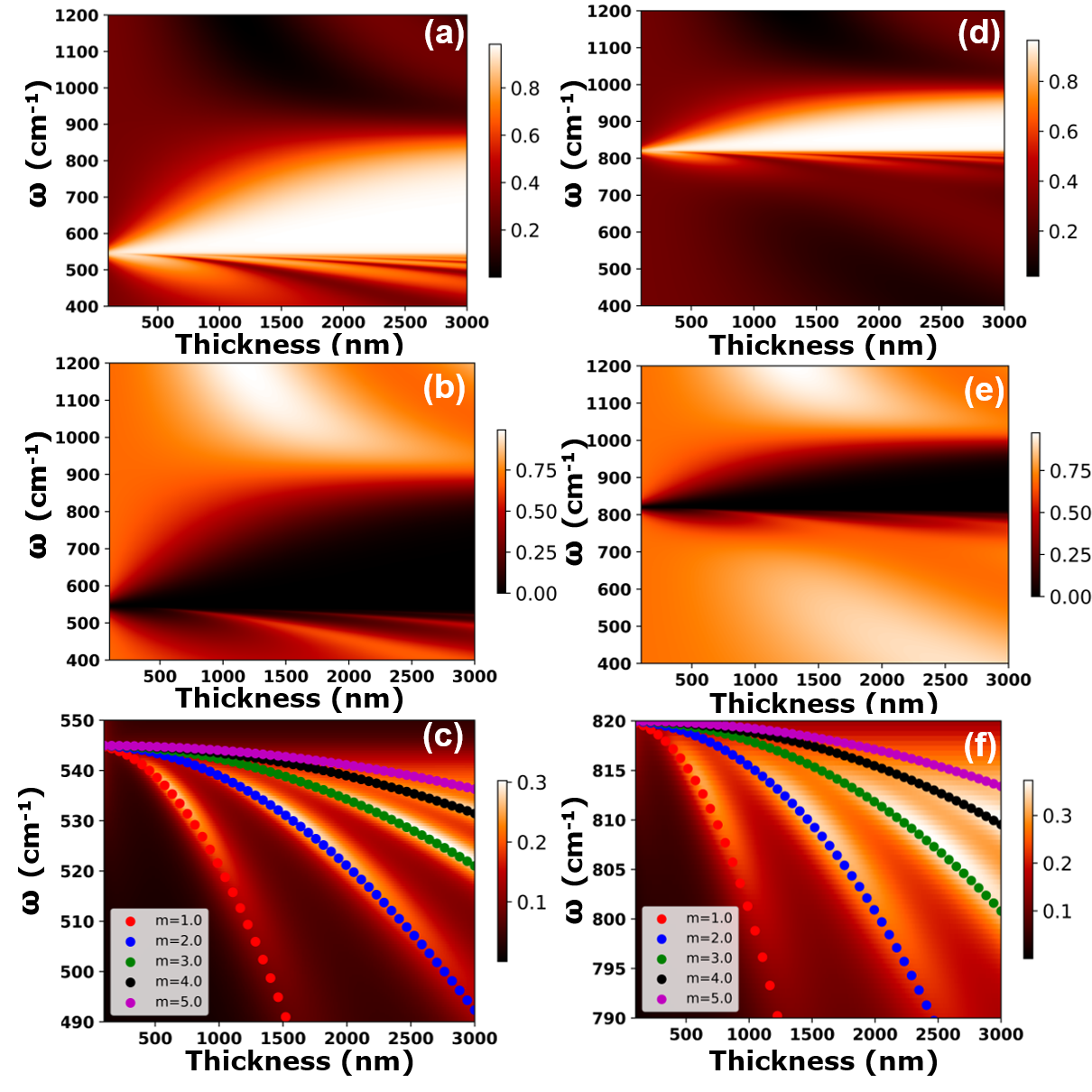}
    \caption{(a)-(c) correspond to the reflectance, transmittance, and absorbance, respectively, of $\alpha$-MoO$_{3}$ for $y-$ polarized light as a function of thickness and frequency. (d)-(f) represent the reflectance, transmittance, and absorbance, respectively, of $\alpha$-MoO$_{3}$ for $x-$ polarized light as a function of thickness and frequency. The scatter plots in absorption color plots correspond to Fabry Perot modes of $\alpha$-MoO$_{3}$ film.}
\end{figure}

Thickness of the hyperbolic vdW crystal is another vital parameter that influences the optical response. We investigate the optical responses as a function of the frequency and thickness, as shown in Fig.~2. Reflectance of $x$- and $y$-polarized light, with the frequencies greater than TO phonons, increases with the increasing thickness (Fig.~2(a)$\&$(d)). This is consistent with increased reflectance from a thicker metallic film. Contrary to reflectance, the transmittance of $x$- and $y$-polarized light, with frequencies greater than TO phonons, decreases with the increasing thickness (Fig.~2(b)$\&$(e)). Furthermore, several sharp dips in the reflectance color plot are observed, for both $x$- and $y$-polarized light, below their respective TO phonon frequencies (Fig.S2 of supporting information). At the frequency of dips in the reflectance color plot, strong peaks are observed in the absorbance and transmittance color plot. This feature is more clearly visible in the absorbance color plot shown in Fig.~2(c)$\&$(f). Since dielectric permittivity of $\alpha$-MoO$_{3}$ just below the TO phonon frequency is significantly large, the thin film of $\alpha$-MoO$_{3}$ works as a Fabry-Perot cavity in the mid-IR spectral region resulting in the observed discrete absorption modes. An analytical model in Eq.~2 suggests (see Sec. S2 of supporting information) the relation between frequency, thickness, and order of the Fabry Perot mode. In Eq.~2, $K^j = \frac{1}{\varepsilon^j_\infty} (\frac{m\pi c}{d})^2$, $m$ is order of the mode, $c$ is speed of light in free space and $d$ the thickness of $\alpha$-MoO$_{3}$ film. The analytical prediction from Eq.~2, as shown in Fig.~2(c)$\&$(f) for the different absorption modes of Fabry-Perot cavity, are in excellent agreement with the results from our TMM model:
\begin{equation}
    \omega^2 = \frac{((\omega^j_{LO})^2 + K^j) - \sqrt{((\omega^j_{LO})^2 + K^j)^2 - 4K^j((\omega^j_{TO})^2)}}{2}
\end{equation}
\begin{figure}
    \centering
    \includegraphics[width=\columnwidth]{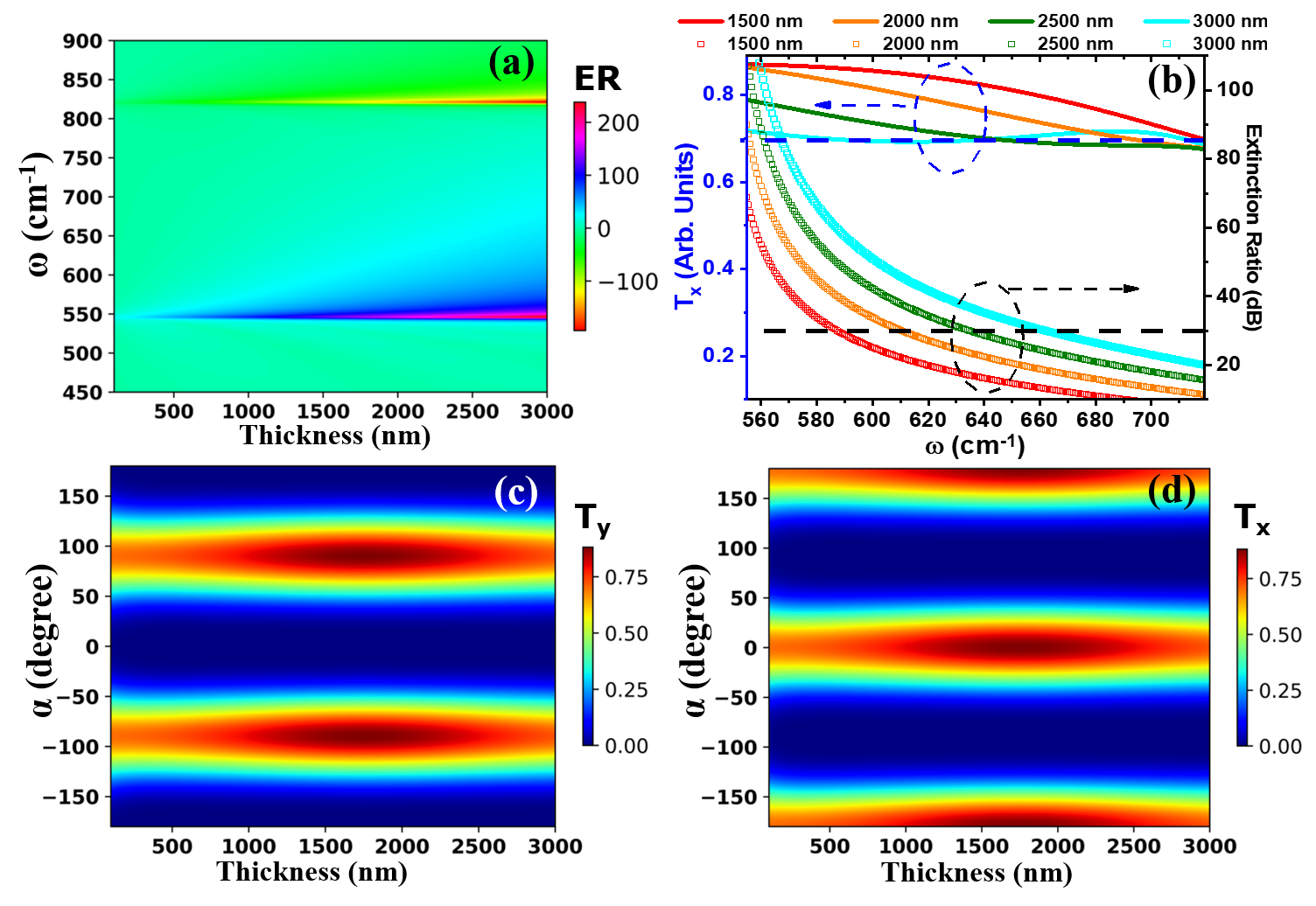}
    \caption{(a) Extinction ratio (in dB) of $\alpha$-MoO$_{3}$ as a function of thickness and frequency.(b) Figure of merits (i.e. Transmission efficiency and Extinction Ratio) of $\alpha$-MoO$_{3}$ based polarizer in RS band 1 at different thicknesses. (c)-(d) corresponds to the transmittance of $x-$ and $y-$ polarized light respectively at a frequency of 546 cm $^{-1}$ as a function of thickness and $\alpha$ -- orientation of the optical axis of $\alpha$-MoO$_{3}$ with respect to the polarization plane of linearly polarized light.}
\end{figure}

With the above-given understanding of the polarization-dependent optical responses of $\alpha$-MoO$_{3}$ films, we first explore their potential for application to mid-IR polarizers. To characterize such a polarizer, we define two well-known figures of merit -- transmission efficiency and extinction ratio (ER), the latter being given by the formulae:
\begin{align}
    \text{ER} = 10\log(T_x/T_y) 
\end{align}
Here, $T{_x}$ and $T{_y}$ are transmittances of $x$-polarized and $y$-polarized light. The ER at 546 cm$^{-1}$ and 820 cm$^{-1}$ for 100 nm thin film of $\alpha$-MoO$_{3}$ are found to be around 24 dB and 19 dB respectively which is remarkable for a 100 nm thin-film based polarizer. The figures of merit (i.e., ER and transmission efficiency) for $\alpha$-MoO$_{3}$ based polarizer as a function of thickness are shown in Fig.~3(a)-(b) respectively. For the thickness of few microns (1 $\mu$m -3 $\mu$m), we observe an enormous ER of around 200 dB and 190 dB at the frequency of 546 cm$^{-1}$ and 820 cm$^{-1}$ respectively which is ascribed to strong in-plane anisotropy of $\alpha$-MoO$_{3}$ thin films\cite{Ma2018}. Next, we estimate the frequency bandwidth of the polarizer for various thicknesses of $\alpha$-MoO$_{3}$ in RS band-1 by considering transmission efficiency and ER thresholds of 70\% and several tens of dB respectively as shown in Fig.~3(b). A 3000 nm thin $\alpha$-MoO$_{3}$ film based polarizer exhibits a frequency bandwidth of 117 cm$^{-1}$ (~3.24 $\mu$m), 80 cm$^{-1}$ (~2.34 $\mu$m), and 56 cm$^{-1}$ (~1.71 $\mu$m) from 545 cm$^{-1}$ at ER thresholds of 30 dB, 40 dB and 50 dB respectively (shown in Fig.~3(b)) with more than 70$\%$ transmittance. The performance metrics of $\alpha$-MoO$_{3}$ based polarizer at other thicknesses are tabulated in the supporting information. Furthermore, the transmittance of $x$- and $y$- polarized light at 546 cm$^{-1}$ is calculated as a function of polarization angle ($\alpha$), which depicts the transmittance of linearly polarized light through parallel and perpendicular to the polarization axis of the crystal and demonstrates the Malus' law\cite{Zhou2018}. This study suggests the optimized range of thickness for an efficient mid-IR polarizer is 1500 nm to 2500 nm. These studies prove that one can design excellent mid-IR polarizer from $\alpha$-MoO$_{3}$ thin films in RS band 1. Our study further reveals poor transmission efficiency of mid-IR polarizer from $\alpha$-MoO$_{3}$ thin films in RS band 2 (see Fig.~S3 in supporting information) limiting its operational region to RS band 1 only.\par
\begin{figure}
    \centering
    \includegraphics[width=\columnwidth]{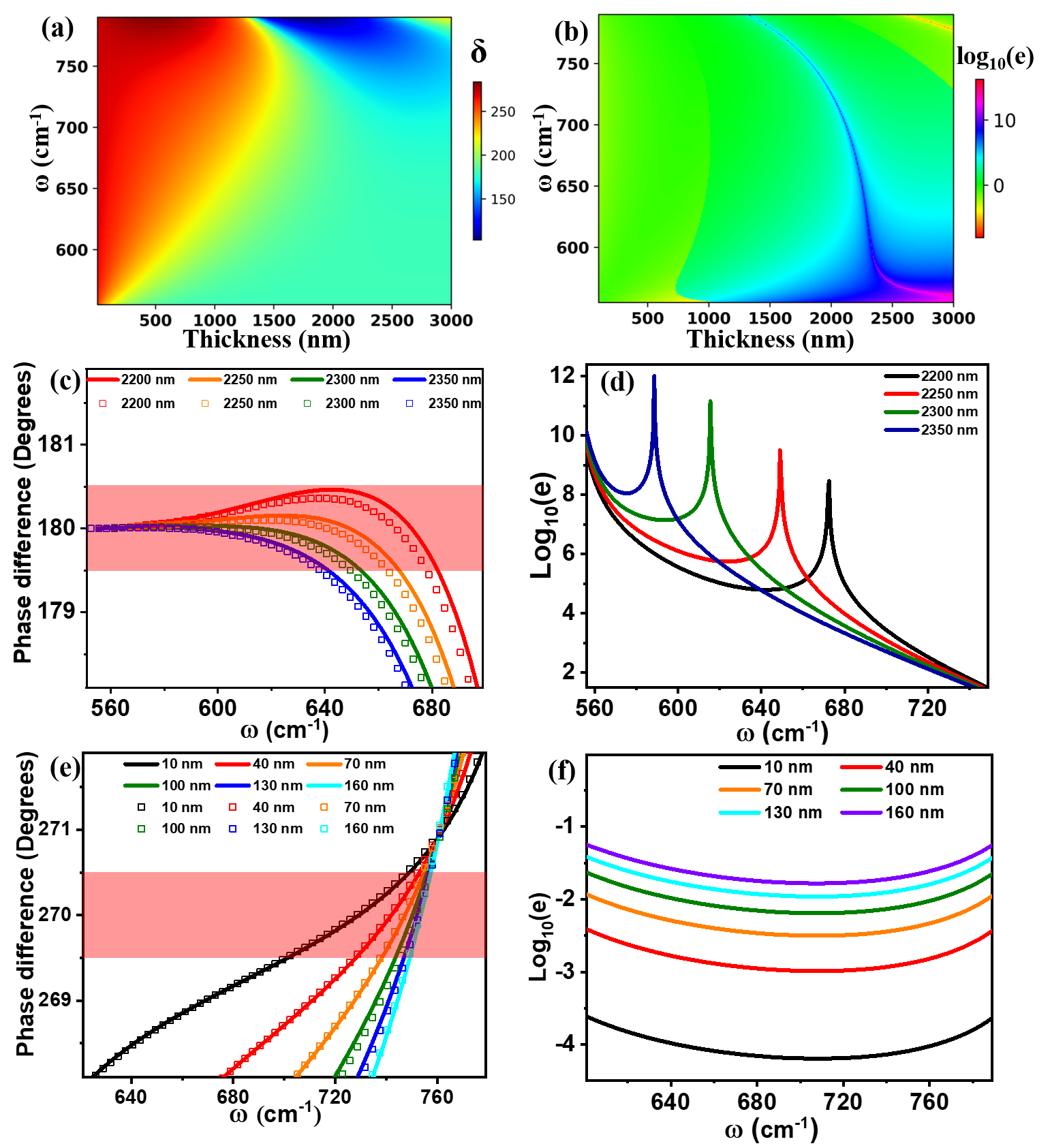}
    \caption{(a) shows the phase difference($\delta$) between $x-$ and $-y$ component of transmitted light due to incidence of linearly polarized light and (b) corresponding ellipticity($e$) as a function of thickness and frequency. (c)-(d) corresponds to phase difference and ellipticity, respectively, at the optimum thicknesses for half-wave plate application. (e)-(f) represents the phase difference and the ellipticity, respectively, at the optimum thicknesses  for quarter-wave plate applications.}
\end{figure}

Next, we explore the prospect of $\alpha$-MoO$_{3}$ thin films for mid-IR wave-plate applications. Fig.~4(a) shows that $\alpha$-MoO$_{3}$ thin film can produce a phase difference of $\pi$/2 and $\pi$ between both components of transmitted light which is a fundamental requirement for a quarter waveplate (QWP)and half wave-plate(HWP). However, the ellipticity color plot in Fig.~4(b) suggests elliptical and linearly polarized transmitted light from $\alpha$-MoO$_{3}$ thin film. From the color plot, the optimum thickness range for the HWPs is found around 2200 nm to 2400 nm, where from the ellipticity of transmitted light (shown in Fig.~4(b)) we ascertain that we obtain linearly polarized output. Ellipticity is evaluated as a function of thickness and frequency using the following relation \cite{Yang2017,trager2012springer}: 
\begin{equation}
    e = \frac{1 + s\sqrt{1 - 4\sin^2{\delta}\cdot\frac{t_x^2  t_y^2}{(t_x^2 + t_y^2)^2}}}{1 - s\sqrt{1 - 4\sin^2{\delta}\cdot\frac{t_x^2  t_y^2}{(t_x^2 + t_y^2)^2}}}
\end{equation} 
where $s = -1$ when $t_y^2 - t_x^2 < 0$ otherwise $s = 1$. $t_x$ and $t_y$ are absolute value of transmission coefficient of $x-$ and $y-$ component of transmitted light. Here $\delta = \delta_y - \delta_x$ represents the phase difference between $y$ and $x$ component of transmitted light. With this definition, the logarithmic value of ellipticity has been considered to investigate the properties of transmitted light. Fig.~4(c) reveals the achromatic phase retardation property-- phase retardation is relatively independent of wavelength (frequency) -- of $\alpha$-MoO$_{3}$ thin film at the thickness of 2200 nm and 2350 nm. The frequency bandwidth for achromatic HWPs with the retardation tolerance of $\pm 0.5^\circ$ in phase difference is found around 105 cm$^{-1}$ (~3.86 $\mu$m) (from 555 cm$^{-1}$ to 660 cm$^{-1}$). This result also unveils another exotic property of $\alpha$-MoO$_{3}$ -- achromatic retardation-- which is generally attained using combination of multiple waveplates\cite{Messaadi2018,Pisano2006,Stoyanova2019}. We further validate the phase difference results from the TMM model with FDFD simulation (shown as scatter plot in Fig~4(c)), which is found to be in good agreement. The large values of $\log(e)$ shown in Fig.~4(d), indicate almost linearly polarized output. The peaks in the ellipticity spectra correspond to the frequency at which the phase difference is exactly $\pi$. At the phase difference of $\pi$, major axis of polarization ellipse (represented by denominator of Eq.~4) become negligible, resulting in the large value of ellipticity which are observed as a peak in the ellipticity spectra for different thicknesses of $\alpha$-MoO$_{3}$.

For application to QWP design, we found that a 10 nm thin $\alpha$-MoO$_{3}$ film can produce a phase difference of $90^\circ \pm0.5^\circ$ with a broad spectral range from 700 cm$^{-1}$ to 750 cm$^{-1}$. However, the fabrication of such small thickness of single crystal $\alpha$-MoO$_{3}$ might be difficult using traditional vapor deposition approaches \cite{Wang2017}. The bandwidth of QWP decreases with increasing thickness of $\alpha$-MoO$_{3}$ as shown in Fig.~4(e). The scatter plots in Fig.~4(e) correspond to the FDFD numerical simulation results for phase difference substantiating the results of our TMM model. Furthermore, ellipticities of transmitted light for the various thicknesses are found smaller than 1, indicating the elliptically polarized transmitted light. Moreover, it has been observed that $\alpha$-MoO$_{3}$ does not possess adequate phase difference in RS band 2 for the application of HWPs and QWPs (See Fig.~S3 in supporting information) and therefore, its operational spectral range is limited to RS band 1 only. These studies reveal the potential of $\alpha$-MoO$_{3}$ thin films for application to phase retarders. A 2300 nm $\alpha$-MoO$_{3}$ film acts like half waveplate over a broadband of 555 cm$^{-1}$ to 660 cm$^{-1}$ ($\sim 2.86 \mu$m) and 100 nm thin film acts as a elliptical polarizer from 743 cm$^{-1}$ to 758 cm$^{-1}$ ( $\sim 260$ nm) frequency range.

In conclusion, our theoretical investigations of the optical responses suggest that the recently explored naturally hyperbolic material -- $\alpha$-MoO$_{3}$ -- exhibits several exotic phenomena due to strong in-plane birefringence and hyperbolic nature in the mid-IR spectral region of light. We show that these properties can be harnessed to design optical devices such as mid-IR polarizers and phase retarders with sub-wavelength thickness. We presented the design of a high performance $\alpha$-MoO$_{3}$ based mid-IR polarizer with an ER of more than 30 dB and transmission efficiency of 70\%, without the need for lithographic patterning. We also optimized thin-film of $\alpha$-MoO$_{3}$ to function as an achromatic half waveplate and an elliptical quarter waveplate in the bandwidth of 555 cm$^{-1}$ to 660 cm$^{-1}$ and 740 cm$^{-1}$ to 760 cm$^{-1}$ respectively. Our analysis and optimization reveals the potential of $\alpha$-MoO$_{3}$ as a lithography free alternative for miniaturized mid-IR optical components.
\section{Acknowledgement}
A.K. acknowledges funding from the Department of Science and Technology grant numbers SB/S2/RJN-110/2017, ECR/2018/001485 and DST/NM/NS-2018/49. S.D. acknowledges financial support from Institute Postdoctoral Fellowship IIT Bombay. N.R.S. acknowledges financial support from Council of Scientific $\&$ Industrial Research fellowship No: 09/087(0997)/2019-EMR-I.
\bibliography{Main}
\end{document}